\documentclass[conference]{IEEEtran}
\IEEEoverridecommandlockouts
\usepackage{amsmath,amssymb,amsfonts,amsthm}
\usepackage{algorithmic}
\usepackage[linesnumbered,ruled,vlined]{algorithm2e} 

\usepackage{comment}
\usepackage{cancel}
\usepackage[export]{adjustbox}
\usepackage{ragged2e} 
\usepackage[utf8]{inputenc} 
\usepackage{standalone}
\usepackage{pgfplots}
\usepackage{lipsum} 
\usepackage{setspace}
\usepackage{multirow}
\usepackage{tabularray}

\usepackage{tikz}
\usetikzlibrary{shapes,arrows}
\usetikzlibrary{positioning}
\usetikzlibrary{external}
\usetikzlibrary{patterns}
\usepackage{pstricks}
\usepackage{float}
\usepackage[caption=false,font=normalsize,labelfont=sf,textfont=sf]{subfig}
\usepackage{graphicx}
\usepackage{textcomp}
\usepackage{xcolor}
\usepackage[ruled,vlined]{algorithm2e}
\def\BibTeX{{\rm B\kern-.05em{\sc i\kern-.025em b}\kern-.08em
    T\kern-.1667em\lower.7ex\hbox{E}\kern-.125emX}}

\usepackage{booktabs}
\usepackage{textcomp}
\usepackage{stfloats}
\usepackage{url}
\usepackage{verbatim}
\usepackage{graphicx}
\usepackage{cite}

\usepackage{array}
\usepackage{tabularray}

\newcommand{\bG}{{\bf G}}

\newcommand{\bu}{{\bf u}}
\newcommand{\bv}{{\bf v}}
\newcommand{\bx}{{\bf x}}

\newcommand{\bp}{{\bf p}}
\newcommand{\bs}{{\bf s}}
\newcommand{\bg}{{\bf g}}

\newcommand{\bui}[1]{\bu_{#1}}

\newcommand{\myRectangle}[6]{% #1 = x, #2 = y, #3 = largeur, #4 = hauteur, #5 = couleur de ligne, #6 = couleur de remplissage
  \draw[#5, fill=#6] (#1,#2) rectangle (#1+#3,#2+#4);
}

\newtheorem{thm}{Theorem}[section]

\newtheorem{example}[thm]{Example}

\DeclareMathOperator*{\biggcup}{\bigcup}
\DeclareMathOperator*{\summ}{\sum}
\DeclareMathOperator*{\prodd}{\prod}

\def\BibTeX{{\rm B\kern-.05em{\sc i\kern-.025em b}\kern-.08em
    T\kern-.1667em\lower.7ex\hbox{E}\kern-.125emX}}

\begin{document}

\title{Computing the Low-Weight codewords of Punctured and Shortened Pre-Transformed polar Codes}
\author{%

      \IEEEauthorblockN{Malek Ellouze, Romain Tajan, Camille Leroux and Christophe Jégo}
  \IEEEauthorblockA{University of Bordeaux \\
  Bordeaux INP - ENSEIRB-MATMECA \\
  IMS Lab, UMR CNRS 5218, France \\
                    Email: \{surname.name\}@ims-bordeaux.fr}
    \and
  \IEEEauthorblockN{Charly Poulliat}
  \IEEEauthorblockA{University of Toulouse \\
  INPT-ENSEEIHT \\
  IRIT Lab, UMR CNRS 5505, France \\
                    Email: charly.poulliat@enseeiht.fr}

}

\maketitle

\begin{abstract}
In this paper, we present a deterministic algorithm to count the low-weight codewords of punctured and shortened pure and pre-transformed polar codes. The method first evaluates the weight properties of punctured/shortened polar cosets. Then, a method that discards the cosets that have no impact on the computation of the low-weight codewords is introduced. A key advantage of this method is its applicability, regardless of the frozen bit set, puncturing/shortening pattern, or pre-transformation. Results confirm the method's efficiency while showing reduced computational complexity compared to state-of-the-art algorithms.
\end{abstract}

\section{Introduction}

The growing interest regarding polar codes \cite{b1} is due to their ability to achieve the channel capacity asymptotically. Polar codes, nonetheless, possess two primary constraints. The first limitation arises from the fact that, for moderate code lengths, polar codes have poor distance properties. 
The enhancement of the distance properties of polar codes saw a notable stride with the introduction of precoding. Among the precoding techniques, the concatenation of cyclic redundancy check (CRC) with polar codes \cite{b3} combined with a successive cancellation list (SCL) decoder \cite{b2} helped improve the distance properties and the decoding performance of polar codes. A more recent alternative is to apply precoding before the polar transformation, as seen in \textit{Dynamic Frozen Bits (DFB) } polar codes  \cite{b14} and \textit{Polarization-Adjusted Convolutional} (PAC) codes  \cite{b13}.

The second limitation is that from the pure polar code construction perspective, polar codes can only have sizes expressed as powers of $2$. Methods such as puncturing and shortening have been introduced to manage the constraint on code sizes, enabling the construction of rate-compatible polar codes and leading to the development of various puncturing and shortening strategies \cite{b36, b27, b26}. 

% Several puncturing strategies have been introduced to enhance the performance of punctured polar codes, such as the Quasi-uniform Puncturing (QUP) algorithm introduced in \cite{b36} and the bit-reversal scheme \cite{b27}.
%  For shortened polar codes, numerous shortening patterns were proposed to improve their performance. \cite{b26} first introduced the shortening of polar codes as a new puncturing scheme based on the column weights of the generator matrix. \cite{b27} proposed a bit reversal approach for the puncturing and shortening patterns, proving better performance.
Shortening is, however, more challenging in the case of PAC and DFB codes, as the values fixed by the shortening patterns can be modified due to the precoding process. To solve this issue, \cite{b244} suggested a constraint on the convolutional precoding for PAC codes that enables preserving the shortening pattern initial values. \\
All of the aforementioned strides were coupled with the introduction of methods to evaluate the distance properties of pure, precoded, rate-compatible polar codes as it enables estimating the performance of codes under Maximum-Likelihood (ML) decoding.
The minimum distance computation for polar codes originated in \cite{b4}. In \cite{b5}, the computation of the number of minimum weight codewords assumes frozen sets of decreasing monomial codes. A more general yet complex and non-deterministic approach given in \cite{b3} estimates partial distances using Monte Carlo simulations with large list sizes. In \cite{b6}, a deterministic algorithm for calculating polar code weight distribution was proposed, but its high computational complexity restricted its use to codes up to length 128 under certain conditions. \cite{li2021} introduced a method for computing the average partial weight spectrum of pre-transformed polar codes. \cite{rowshan2023} and \cite{rowshan2023closed} focused on partial weight spectrum calculations but were limited to decreasing monomial constructions and are not applicable to pre-transformed polar codes. \cite{b17} and \cite{zunker2024} developed low-complexity methods for computing the minimum distance properties of pre-transformed polar codes, but only when the minimum distance matches that of pure polar codes.
In \cite{powerful}, low-weight codewords are enumerated for polar, rate-compatible, and precoded polar codes through a recursive decomposition. However, it can yield high complexity in the case of precoded and rate-compatible polar codes. \\
This paper introduces a new low-complexity algorithm for the computation of the partial spectrum for rate-compatible pure and pre-transformed polar codes. Unlike \cite{li2021}, this method is deterministic, i.e it allows the determination of the exact reduced spectrum. In contrast to \cite{rowshan2023, rowshan2023closed, b17, zunker2024}, it can be adapted to punctured/shortened polar codes and compute the partial spectrum beyond just the minimum distance properties. This approach is an extension to the one presented in \cite{b23} and offers the advantage of not assuming any specific structure (a) for the frozen bit set, (b) for the pre-transformation or (c) for the shortening and puncturing patterns. It, therefore, aids in the design of puncturing shortening patterns, pre-transformation parameters, and frozen bit sets for punctured/shortened polar codes, thereby enhancing their decoding performance. 
\section{Preliminaries}
\label{sec:PolarCoDec}

% \subsection{Definitions}
% In the following, we note ${\mathbf{a}}_{i}^{j} \triangleq (a_i, a_{i+1}, … a_j)$ and
%  $\mathbb{F}_2=\{0,1\}$ the finite field  of order two. Let $\mathbb{F}_2^N$ the vector space of all $N$-vectors/$N$-tuples ${\mathbf{x}}_{0}^{N-1}$ defined over $\mathbb{F}_2$, endowed with its usual vector addition noted $\oplus$.
% %The Hamming weight of a vector $\bf{c}_{0}^{N-1} \in \mathbb{F}_2^{N}$, defined as the number of its nonzero components, is noted $w(\bf{c}_{0}^{N-1})$.
% Let $\mathcal{C}$ be a $(N,K)$ linear block code of length $N$ and dimension $K$ and of code rate $R=K/N$. Its $2^K$ codewords form  a $K$-dimensional vector subspace of $\mathbb{F}_2^N$.  The Hamming weight of a codeword $\mathbf{c}\in \mathbb{F}_2^{N}$, defined as the number of its nonzero components, is noted $w(\mathbf{c})$.
% %The Hamming distance $d$ between two vectors $\bf{c}_{0}^{N-1}$ and $\bf{\check{c}}_{0}^{N-1}$ is defined as the number of indexes for which they differ.
% The Hamming distance $d(\mathbf{c},\check{\mathbf{c}})=w(\mathbf{c}\oplus\check{\mathbf{c}})$ between two codewords $\mathbf{c}$ and $\mathbf{\check{c}}$ is defined as the number of components for which they differ. The minimum distance of the code $\mathcal{C}$ is given by

% %$\mathcal{C}$ is called a linear code characterized by $(N,K,d^*)$ where $d^*$ denotes the minimal distance of the code : 
% \begin{equation}
%     d^*(\mathcal{C}) = \min_{\mathbf{c} \in \mathcal{C}, \mathbf{c} \neq \mathbf{0}} w(\mathbf{c})
% \end{equation}

\subsection{Polar codes and polar cosets}
The polar code $(N = 2^n,K)$ transformation matrix is given by the $n$-fold polar Kronecker matrix $\bG = {G_2}^{\bigotimes n}$  where  ${(\cdot)}^{\bigotimes n}$ denotes the $n^{th}$ Kronecker product power and $G_2 = 
\begin{bmatrix}
 1 & 0\\ 
 1& 1
\end{bmatrix}$
The codewords of a polar code are obtained such that $\mathbf{x} = \mathbf{u} \bG$, where $\mathbf{u}\in \mathbb{F}_2^N$ is an information vector for which $K$ positions are assigned to the information bits, whereas the remaining ones are \textit{frozen}, i.e set to some known values. This operation is called \textit{rate-profiling}.  We note $\mathcal{F}$ the set of indices of the components of $\mathbf{u}$ corresponding to the frozen bits. 
We refer to polar codes with all frozen bits set to zero as \textit{pure polar codes}.

\subsection{Pre-Transformed polar codes}
A pre-transformation of polar codes consists in applying a linear mapping before the multiplication with the transformation matrix. This can be conveyed as a vector-matrix multiplication with an upper triangular matrix $T$. The overall encoding process can be described as $ \bx = \underset{\bu}{\underbrace{\bv T}} \bG $. Various polar code variants, such as DFB and PAC codes, can be unified under the concept of pre-transformed polar codes.\\
In the case of PAC codes \cite{b13}, the pre-transformation consists of a convolutional encoding using the generator function $\bg$ of degree m with coefficients $[g_0, g_1, ..., g_{m-1}]$, i.e, given a vector $v_i$, the associated $u_i$ is obtained as: $u_i = \bg(\bv_0^i) = \sum_{j=0}^{m-1}g_j v_{i-j}$.
\subsection{Rate-compatible polar codes}
Let $\mathcal{S}$ and $\mathcal{P}$ denote the shortening and puncturing patterns, respectively. We also note $S = |\mathcal{S}|$ and $P = |\mathcal{P}|$. The lengths of a shortened and punctured polar code, respectively given by $N_s = N-S$ and $N_p = N-P$, are derived from a parent polar code of length $N$. When shortening a polar code, a designated subset of the parent code is selected. Within this subset, a total of $S$ codeword bits are fixed to a predetermined value, e.g 0. Since the shortened codeword bits are perfectly known to the decoder, they lead to very reliable elements $\mathcal{U}_{\mathcal{S}}$ of $\bu$ qualified as \textit{overcapable} \cite{b27} and need to be frozen. We denote by $\mathcal{F}_{\mathcal{S}} = \mathcal{F} \bigcup {\mathcal{U}_{\mathcal{S}}}$ the set of frozen bits in the case of shortened polar codes.  \\
% From the decoder's perspective, since the shortening pattern bits are known, i.e $\bx_{\mathcal{S}} = \bf{0}$, the LLRs corresponding to those values are initialised with a large positive value. The corresponding bits are considered frozen. \\
In the puncturing process of polar codes, a total of $P$ codeword bits are treated as erased and, consequently not transmitted. The unreliability of codeword bits $\bx_{\mathcal{P}}$, due to being unknown to the decoder, affects the initially transmitted vector $\bu$. As a result, a set $\mathcal{U}_{\mathcal{P}}$ of  bits of $\bu$ are deemed \textit{incapable} \cite{b27}, and must be frozen. The set of frozen bits for punctured polar codes is denoted as $\mathcal{F}_{\mathcal{P}} = \mathcal{F} \bigcup \mathcal{U}_{\mathcal{P}}$.\\
% From the decoder's standpoint, the punctured bits are treated as erased, leading to their associated LLRs being set to zero.
% \cite{b27} introduced puncturing and shortening patterns based on bit-reversal permutation. Given $\bf{B_N}$, the bit-reversal permutation for bits coordinates of length N, the shortening pattern is defined as $\mathcal{S} = \bf{B_N}$$(N_s, ..., N-1)$ and the puncturing pattern as $\mathcal{P} = \bf{B_N}$$(0, ..., N_p-1)$. Unless specified otherwise, the shortening patterns used in this paper are the ones defined with the bit-reversal permutation. \\
In the context of pre-transformed polar codes, for punctured polar codes, the pre-transformation remains unaffected as the punctured codewords bits are not transmitted. It, therefore, does not alter the pre-transformation rules used for pure polar codes.
However, in the case of shortening pre-transformed polar codes, \cite{b244} introduced a constraint on the pre-transformation. This constraint is designed to ensure that the shortening condition is met, i.e. $\bx_{\mathcal{S}} = \bf{0}$. In the case of PAC codes, the constrained pre-transformation consists in: 
\begin{equation}
    u_i =  \left\{\begin{matrix}
 0 \quad \text{if} \quad i \in \mathcal{S}\\
 \sum_{j=0}^{m-1}g_j v_{i-j} \quad \text{otherwise}
\end{matrix}\right.
\label{eq_shortened_pac}
\end{equation}
In this paper, in the case of shortened pre-transformed polar codes, we consider the constraint on the pre-transformation in Equation \eqref{eq_shortened_pac}.
\section{Minimum Weight Enumeration Function (MWEF) and Reduced Weight Enumeration Function (RWEF) of rate-compatible polar cosets}

\label{sec_min_weight}
\subsection{Computation of the MWEF and RWEF of polar cosets}
\label{w_general}
As in \cite{b6}, given $\bu_{0}^{i-1}  \triangleq (u_0, u_{1}, … ,u_{i-1}) \in \mathbb{F}_2^{i-1}$ and $u_i \in \mathbb{F}_2 $, a polar coset ${\mathcal{C}_N}$ can be defined as:   
\begin{equation}
    {\mathcal{C}_N}(\bu_{0}^{i})=  \{ [\bu_{0}^{i},\bu_{i+1}^{N-1}]\bG | \bu_{i+1}^{N-1} \in \mathbb{F}_2^{N-i-1}  \}
\end{equation}
A polar coset ${\mathcal{C}_N}(\bu_{0}^{i})$ thus describes the codewords' affine space generated by the prefix $\bu_{0}^{i}$. and can also be expressed as: 
\begin{equation}
    {\mathcal{C}_N}(\bu_{0}^{i})=  \bp \oplus \left\{ \bs \in \mathbb{F}^n_2 \middle|   \bs H^{(i)T}= 0  \right\}
\end{equation}
where $ \bp= \bui{0}^{i-1} \bG_{0}^{i-1}$ and $H^{(i)}$ denotes a parity check matrix associated to $\bG_{i+1}^{N-1}$ the $(N-i-1)$ last rows of $\bG$, i.e. $\bG_{i+1}^{N-1} H^{(i)T} = 0$. \\
As shown in \cite{b19}, in the specific case of polar codes with Arikan's kernel, there exists an extended code $[\bs, \textbf{t}, u_i]$ associated to an extended parity check matrix whose Tanner graph is a tree. Therefore, the message passing formalism is used to compute the distance properties of polar cosets. In \cite{b6}, the Minimum Weight Enumeration (WEF) function $A_N( {\mathcal{C}_N}(\bu_{0}^{i}))(X)$ is defined as: 
\begin{equation}
    A_N( {\mathcal{C}_N}(\bu_{0}^{i}))(X) \triangleq \sum_{w = 0}^{N} A_w X^w
\end{equation}
where $A_w$ is the number of words of ${\mathcal{C}_N}(\bu_{0}^{i})$ with weight $w$. In this work, we define the MWEF $A_N^{*}( {\mathcal{C}_N}(\bu_{0}^{i}))(X)$ and RWEF $A_N^{w_{end}}( {\mathcal{C}_N}(\bu_{0}^{i}))(X)$ respectively as: 
\begin{equation}
    \left\{\begin{matrix}
 A_N^{*}( {\mathcal{C}_N}(\bu_{0}^{i}))(X) \triangleq  A^* X^{w^*}
 \\
A_N^{w_{end}}( {\mathcal{C}_N}(\bu_{0}^{i}))(X) \triangleq \sum\limits_{w = 0}^{w_{end}} A_w X^w
\end{matrix}\right.
\end{equation}
In short, the RWEF considers the monomials associated to a weight up to $w_{end}$. MWEF is a specific case of RWEF where $w_{end} = w^*$ where $w^*$ is the minimum weight of the considered coset.
In \cite{b6}, message passing rules are developed in order to compute the WEF of a polar coset. Those rules are adapted in the following in order to compute the RWEF or more specifically the MWEF of a polar coset. We present in the following the message passing rules for the computation of the RWEF.

% For each message coming from a variable or parity node, a vector   $\boldsymbol{\theta}_{x}  = \begin{pmatrix}
%  {\theta}_{x}^{(0)}   \\
%  {\theta}_{x}^{(1)} 
% \end{pmatrix}$ is associated, where ${\theta}_{x}^{(0)}$ and ${\theta}_{x}^{(1)}$ denote the RWEF of the configurations for $x=0$ and $x=1$, respectively. \\

\begin{figure}[htbp]
  \centering
  \subfloat[][Parity node case]
  {
    \scalebox{0.38}{\pagecolor{white}

\begin{tikzpicture} 
\path[use as bounding box] (2, -4) rectangle (10, 11);
% Style definitions
\tikzset{variable_node_style/.style={regular polygon,regular polygon sides=4,draw=green!40!black,fill=green!10!white, inner sep = 3pt}};
\tikzset{check_node_style/.style={circle,draw=blue!50!black,fill=blue!10!white, inner sep = 3pt}};

\tikzset{variable_node_style1/.style={regular polygon,regular polygon sides=4,draw=green,fill=green!10!white, inner sep = 3pt}};
\tikzset{check_node_style1/.style={circle,draw=blue!50!black,fill=blue!10!white, inner sep = 3pt}};
\tikzset{edge_style/.style={draw=black,line width=1pt }};

\draw [dashed] (4.1,3.5) -- (5.9,3.5) -- (5.9,1.5)--(4.1,1.5) -- cycle  ; 
\fill[teal!30!white] (4.1,3.5) -- (5.9,3.5) -- (5.9,1.5)--(4.1,1.5) -- cycle;

\draw [dashed] (6.1,3.5) -- (7.9,3.5) -- (7.9,1.5)--(6.1,1.5) -- cycle;
\fill[lime!30!white] (6.1,3.5) -- (7.9,3.5) -- (7.9,1.5)--(6.1,1.5) -- cycle;

\draw [dashed] (5.1,8.5) -- (6.9,8.5) -- (6.9,10.5)--(5.1,10.5) -- cycle  ;
\fill[pink!30!white] (5.1,8.5) -- (6.9,8.5) -- (6.9,10.5)--(5.1,10.5) -- cycle  ;

\draw [dashed] (4.5,8.2) -- (7.5,8.2) -- (7.5,3.8)--(4.5,3.8) -- cycle  ;
\fill[lightgray!30!white] (4.5,8.2) -- (7.5,8.2) -- (7.5,3.8)--(4.5,3.8) -- cycle  ;

% Variable Nodes

\node [variable_node_style, label=right:{$f$}] (c1) at (6,6) {};

\node [variable_node_style] (c11) at (5.5,3) {};
\node [variable_node_style] (c12) at (4.5,3) {};

\node [variable_node_style] (c21) at (6.5,3) {};
\node [variable_node_style] (c22) at (7.5,3) {};

\node [variable_node_style] (c31) at (5.5,9) {};
\node [variable_node_style] (c32) at (6.5,9) {};

% Check Nodes
\node [check_node_style,label=right:{$x_0$}] (x1) at (5,4) {} ;
\node [check_node_style,label=left:{$x_1$}] (x2) at (7,4) {};
\node [check_node_style,label=right:{$x_2$}] (x3) at (6,8) {};

\node [check_node_style] (x11) at (4.25,2) {};
\node [check_node_style] (x12) at (4.75,2) {};
\node [check_node_style] (x13) at (5.25,2) {};
\node [check_node_style] (x14) at (5.75,2) {};

\node [check_node_style] (x21) at (6.25,2) {};
\node [check_node_style] (x22) at (6.75,2) {};
\node [check_node_style] (x23) at (7.25,2) {};
\node [check_node_style] (x24) at (7.75,2) {};

\node [check_node_style] (x31) at (5.25,10) {};
\node [check_node_style] (x32) at (5.75,10) {};
\node [check_node_style] (x33) at (6.25,10) {};
\node [check_node_style] (x34) at (6.75,10) {};

% Edges
\path [edge_style] (x1) -- (c1);
\path [edge_style] (x2) -- (c1);
\path [edge_style] (x3) -- (c1);

\path [edge_style] (x1) -- (c11);
\path [edge_style] (x1) -- (c12);

\path [edge_style] (x2) -- (c21);
\path [edge_style] (x2) -- (c22);

\path [edge_style] (x3) -- (c31);
\path [edge_style] (x3) -- (c32);

\path [edge_style] (c12) -- (x11);
\path [edge_style] (c12) -- (x12);
\path [edge_style] (c11) -- (x13);
\path [edge_style] (c11) -- (x14);

\path [edge_style] (c21) -- (x21);
\path [edge_style] (c21) -- (x22);
\path [edge_style] (c22) -- (x23);
\path [edge_style] (c22) -- (x24);

\path [edge_style] (c31) -- (x31);
\path [edge_style] (c31) -- (x32);
\path [edge_style] (c32) -- (x33);
\path [edge_style] (c32) -- (x34);

\draw  [->](5,4.25) -- (5.75,5.75);
\draw  [->](7,4.25) -- (6.25,5.75);
\draw  [->, red](6.1,6.25) -- (6.1,7.8);
\node [below] at (4.75,5) {$\boldsymbol{\theta}_{x_0\rightarrow f}$};
\node [below] at (7.25,5) {$\boldsymbol{\theta}_{x_1\rightarrow f}$};
\node [below] at (6.7,7) {\textcolor{red}{$\boldsymbol{\theta}_{f \rightarrow x_2}$}};

\draw  [<-](5.9,6.25) -- (5.9,7.8);
\node [below] at (5.2,7) {\textcolor{black}{$\boldsymbol{\theta}_{x_2\rightarrow f}$}};

\node [below] at (4.1,4) {$T_{x_0}$};
\node [below] at (7.8,4) {$T_{x_1}$};
\node [below] at (5,11) {$T_{x_2}$};

\node [below] at (5.05
,1.8) {  \large $ \cdots $};
\node [below] at (7.05
,1.8) {  \large $ \cdots $};
\node [below] at (6.05
,10.5) {  \large $ \cdots $};
%%%%%%%%%%%%%%%

\begin{scope}[shift={(-6, -7.5)}]
\node [below] at (9.1,6.37) {  \huge $ H = $};
\draw [ very thick] (15.6,8.3) -- (15.75,8.3) -- (15.75,3.7)--(15.6,3.7);
\draw [ very thick] (10.1,8.3) -- (9.95,8.3) -- (9.95,3.7)--(10.1,3.7);

\node [below] at (10.7,8.7) {  $\mathbf{t_{T_{x_0}}}$};
\node [below] at (12.5,8.7) {  $\mathbf{t_{T_{x_1}}}$};
\node [below] at (14.7,8.7) {  $\mathbf{t_{T_{x_2}}}$};
\node [below] at (11.4,8.7) { $x_0$};
\node [below] at (13.3,8.7) {$x_1$};
\node [below] at (15.5,8.7) {$x_2$};

\draw [dashed] (10.05,8.15) -- (11.5,8.15) -- (11.5,7)--(10.05,7) -- cycle;
\fill[teal!30!white] (10.05,8.15) -- (11.5,8.15) -- (11.5,7)--(10.05,7) -- cycle;

\draw [dashed] (11.55,5.5) -- (13.5,5.5) -- (13.5,6.95)--(11.55,6.95) -- cycle;
\fill[lime!30!white] (11.55,5.5) -- (13.5,5.5) -- (13.5,6.95)--(11.55,6.95) -- cycle;

\draw [dashed] (13.55,5.45) -- (15.65,5.45) -- (15.65,4.4)--(13.55,4.4) -- cycle;
\fill[pink!30!white] (13.55,5.45) -- (15.65,5.45) -- (15.65,4.4)--(13.55,4.4) -- cycle;

\draw [dashed] (10.05,3.85) -- (15.65,3.85) -- (15.65,4.35)--(10.05,4.35) -- cycle;
\fill[lightgray!30!white] (10.05,3.85) -- (15.65,3.85) -- (15.65,4.35)--(10.05,4.35) -- cycle;

\draw [dashed] (11.55,8.15) -- (15.65,8.15) -- (15.65,7)--(11.55,7) -- cycle;
\node [below] at (13.5,7.9) {\huge \textbf{0}};

\draw [dashed] (11.5,5.5) -- (10.05,5.5) -- (10.05,6.95)--(11.5,6.95) -- cycle;
\node [below] at (10.8,6.6) {\huge \textbf{0}};
\draw [dashed] (15.65,5.5) -- (13.55,5.5) -- (13.55,6.95)--(15.65,6.95) -- cycle;
\node [below] at (14.7,6.6) {\huge \textbf{0}};

\draw [dashed] (10.05,5.45) -- (13.5,5.45) -- (13.5,4.4)--(10.05,4.4) -- cycle;
\node [below] at (11.8,5.3) {\huge \textbf{0}};

\node [below] at (11.4,4.35) { $1$};
\node [below] at (13.3,4.35) {$1$};
\node [below] at (15.5,4.35) {$1$};

\node [below] at (10.8,4.37) {  \large \bf 0};
\node [below] at (12.5,4.37) {  \large \bf 0};
\node [below] at (14.5,4.37) {  \large \bf 0};

\node [below] at (10.8,7.9) {  \large $ H_{T_{x_0}} $};
\node [below] at (12.6,6.6) {  \large $ H_{T_{x_1}} $};
\node [below] at (14.7,5.2) {  \large $ H_{T_{x_2}} $};

\node [below] at (5.05,1.8) {  \large $ \cdots $};
\node [below] at (7.05,1.8) {  \large $ \cdots $};
\node [below] at (6.05,10.5) {  \large $ \cdots $};
\end{scope}

\end{tikzpicture}
 }
    \label{first_diagram}
  }
  \quad
  \subfloat[][Variable node case]
  {
    \scalebox{0.38}{\pagecolor{white}

\begin{tikzpicture} 
\path[use as bounding box] (2, -4) rectangle (10.7, 11);

\draw [dashed] (4.1,3.5) -- (5.9,3.5) -- (5.9,1.5)--(4.1,1.5) -- cycle  ; 
\fill[teal!30!white] (4.1,3.5) -- (5.9,3.5) -- (5.9,1.5)--(4.1,1.5) -- cycle;

\draw [dashed] (6.1,3.5) -- (7.9,3.5) -- (7.9,1.5)--(6.1,1.5) -- cycle;
\fill[lime!30!white] (6.1,3.5) -- (7.9,3.5) -- (7.9,1.5)--(6.1,1.5) -- cycle;

\draw [dashed] (5.1,8.5) -- (6.9,8.5) -- (6.9,10.5)--(5.1,10.5) -- cycle  ;
\fill[pink!30!white] (5.1,8.5) -- (6.9,8.5) -- (6.9,10.5)--(5.1,10.5) -- cycle  ;

\draw [dashed] (4.5,8.2) -- (7.5,8.2) -- (7.5,3.8)--(4.5,3.8) -- cycle  ;
\fill[lightgray!30!white] (4.5,8.2) -- (7.5,8.2) -- (7.5,3.8)--(4.5,3.8) -- cycle  ;

% Style definitions
\tikzset{variable_node_style/.style={regular polygon,regular polygon sides=4,draw=green!40!black,fill=green!10!white, inner sep = 3pt}};
\tikzset{check_node_style/.style={circle,draw=blue!50!black,fill=blue!30!white, inner sep = 3pt}};
\tikzset{edge_style/.style={draw=black,line width=1pt }};

% Variable Nodes

\node [check_node_style, label=right:{$x$}] (c1) at (6,6) {};

\node [check_node_style] (c11) at (5.5,3) {};
\node [check_node_style] (c12) at (4.5,3) {};

\node [check_node_style] (c21) at (6.5,3) {};
\node [check_node_style] (c22) at (7.5,3) {};

\node [check_node_style] (c31) at (5.5,9) {};
\node [check_node_style] (c32) at (6.5,9) {};

% Check Nodes
\node [variable_node_style,label=right:{$f_0$}] (x1) at (5,4) {};
\node [variable_node_style,label=left:{$f_1$}] (x2) at (7,4) {};
\node [variable_node_style,label=right:{$f_2$}] (x3) at (6,8) {};

\node [variable_node_style] (x11) at (4.25,2) {};
\node [variable_node_style] (x12) at (4.75,2) {};
\node [variable_node_style] (x13) at (5.25,2) {};
\node [variable_node_style] (x14) at (5.75,2) {};

\node [variable_node_style] (x21) at (6.25,2) {};
\node [variable_node_style] (x22) at (6.75,2) {};
\node [variable_node_style] (x23) at (7.25,2) {};
\node [variable_node_style] (x24) at (7.75,2) {};

\node [variable_node_style] (x31) at (5.25,10) {};
\node [variable_node_style] (x32) at (5.75,10) {};
\node [variable_node_style] (x33) at (6.25,10) {};
\node [variable_node_style] (x34) at (6.75,10) {};

% Edges
\path [edge_style] (x1) -- (c1);
\path [edge_style] (x2) -- (c1);
\path [edge_style] (x3) -- (c1);

\path [edge_style] (x1) -- (c11);
\path [edge_style] (x1) -- (c12);

\path [edge_style] (x2) -- (c21);
\path [edge_style] (x2) -- (c22);

\path [edge_style] (x3) -- (c31);
\path [edge_style] (x3) -- (c32);

\path [edge_style] (c12) -- (x11);
\path [edge_style] (c12) -- (x12);
\path [edge_style] (c11) -- (x13);
\path [edge_style] (c11) -- (x14);

\path [edge_style] (c21) -- (x21);
\path [edge_style] (c21) -- (x22);
\path [edge_style] (c22) -- (x23);
\path [edge_style] (c22) -- (x24);

\path [edge_style] (c31) -- (x31);
\path [edge_style] (c31) -- (x32);
\path [edge_style] (c32) -- (x33);
\path [edge_style] (c32) -- (x34);

\draw  [->](5,4.25) -- (5.75,5.75);
\draw  [->](7,4.25) -- (6.25,5.75);
\draw  [<-](5.9,6.25) -- (5.9,7.8);
\node [below] at (4.75,5) {$\boldsymbol{\theta}_{f_0 \rightarrow x}$};
\node [below] at (7.25,5) {$\boldsymbol{\theta}_{f_1 \rightarrow x}$};
\node [below] at (5.2,7) {$\boldsymbol{\theta}_{f_2 \rightarrow x}$};

\draw  [->, color = red](6.1,6.25) -- (6.1,7.8);

\node [below] at (6.7,7) {\textcolor{red}{$\boldsymbol{\theta}_{x \rightarrow f_2}$}};

\node [below] at (4.1,4) {$T_{f_0}$};
\node [below] at (7.8,4) {$T_{f_1}$};
\node [below] at (5,11) {$T_{f_2}$};

\node [below] at (5.05
,1.8) {  \large $ \cdots $};
\node [below] at (7.05
,1.8) {  \large $ \cdots $};
\node [below] at (6.05
,10.5) {  \large $ \cdots $};
%%%%%%%%

%%%%%%%%%%%%%%%
\begin{scope}[shift={(-6, -7.5)}]
\node [below] at (9.1,6.37) {  \huge $ H = $};
\draw [ very thick] (16.3,8.3) -- (16.45,8.3) -- (16.45,4.2)--(16.3,4.2)   ;
\draw [ very thick] (10.1,8.3) -- (9.95,8.3) -- (9.95,4.2)--(10.1,4.2)  ;

 \draw [dashed] (10.05,8.15) -- (11.5,8.15) -- (11.5,7)--(10.05,7) -- cycle  ;
 \fill[teal!30!white] (10.05,8.15) -- (11.5,8.15) -- (11.5,7)--(10.05,7) -- cycle  ;

  \draw [dashed] (11.55,5.5) -- (13.5,5.5) -- (13.5,6.95)--(11.55,6.95) -- cycle  ;
 \fill[lime!30!white] (11.55,5.5) -- (13.5,5.5) -- (13.5,6.95)--(11.55,6.95) -- cycle  ;

\draw [dashed] (13.55,5.45) -- (15.65,5.45) -- (15.65,4.4)--(13.55,4.4) -- cycle  ;
\fill[pink!30!white] (13.55,5.45) -- (15.65,5.45) -- (15.65,4.4)--(13.55,4.4) -- cycle  ;

\draw [dashed] (11.55,8.15) -- (15.65,8.15) -- (15.65,7)--(11.55,7) -- cycle  ;
\node [below] at (13.5,7.9) {\huge \textbf{0}};

\draw [dashed] (11.5,5.5) -- (10.05,5.5) -- (10.05,6.95)--(11.5,6.95) -- cycle  ;
\node [below] at (10.8,6.6) {\huge \textbf{0}};
\draw [dashed] (15.65,5.5) -- (13.55,5.5) -- (13.55,6.95)--(15.65,6.95) -- cycle  ;
\node [below] at (14.7,6.6) {\huge \textbf{0}};

\draw [dashed] (10.05,5.45) -- (13.5,5.45) -- (13.5,4.4)--(10.05,4.4) -- cycle  ;
\node [below] at (11.8,5.3) {\huge \textbf{0}};

\node [below] at (10.8,7.9) {  \large $ H_{T_{f_0}} $};
\node [below] at (12.6,6.6) {  \large $ H_{T_{f_1}} $};
\node [below] at (14.7
,5.2) {  \large $ H_{T_{f_2}} $};

\node [below] at (5.05
,1.8) {  \large $ \cdots $};
\node [below] at (7.05
,1.8) {  \large $ \cdots $};
\node [below] at (6.05
,10.5) {  \large $ \cdots $};

\draw [dashed] (16.3,4.45) -- (15.7,4.45) -- (15.7,8.15)--(16.3,8.15) -- cycle  ;
\fill[lightgray!30!white] (16.3,4.45) -- (15.7,4.45) -- (15.7,8.15)--(16.3,8.15) -- cycle ;

\node [below] at (16
,8.6) {  \large $ x $};

\node [below] at (16
,8) { \Large \textbf{0}};
\node [below] at (16
,7.5) {  $ 1 $};

\node [below] at (16
,6.7) { \Large \textbf{0}};
\node [below] at (16
,6) {  $ 1 $};

\node [below] at (16
,5.5) { \Large \textbf{0}};
\node [below] at (16
,4.85) {  $ 1 $};

\node [below] at (10.7,8.7) {  $\mathbf{t_{T_{f_0}}}$};
\node [below] at (12.5,8.7) {  $\mathbf{t_{T_{f_1}}}$};
\node [below] at (14.7,8.7) {  $\mathbf{t_{T_{f_2}}}$};
\end{scope}

\end{tikzpicture}
 }
    \label{second_diagram}
  }
  \caption{Graph factor configurations}

  \label{fig:graph_factor}
\end{figure}
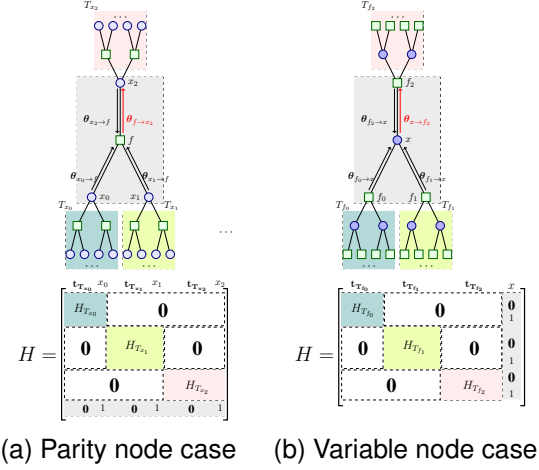

During message passing formalism, two configurations can be encountered. 
The configuration depicted in Figure \ref{first_diagram} shows two variable nodes, $x_0$ and $x_1$, connected to a third variable node $x_2$ via a parity function $f$. The corresponding parity matrix $H$ for the factor graph is also illustrated in the same figure. \\

$T_{x_i} $ is defined as $T_{x_i} = \left\{ \bx | \quad \bx {H^T_{T_{x_i}}} = 0 \right\}$ and to each message coming from a variable node, we associate:
\begin{equation}
    \boldsymbol{\theta}_{x_i \rightarrow f} = \begin{pmatrix}
            \theta^{(0)}_{x_i \rightarrow f} \\
            \theta^{(1)}_{x_i \rightarrow f} 
        \end{pmatrix} = \begin{pmatrix}
            A_N^{w_{end}}(T_{x_i}| x_i = 0)(X) \\
            A_N^{w_{end}}(T_{x_i}| x_i = 1)(X) 
        \end{pmatrix}
\end{equation}

In this case, $\boldsymbol{\theta}_{f \rightarrow x_2}$ can be computed from $\boldsymbol{\theta}_{x_0 \rightarrow f}$ and $\boldsymbol{\theta}_{x_1 \rightarrow f}$ as:
\begin{equation}
    \boldsymbol{\theta}_{f \rightarrow x_2}  = \begin{pmatrix}
       LT_{w_{end}}( {\theta}_{x_0 \rightarrow f}^{(0)} {\theta}_{x_1 \rightarrow f}^{(0)} + {\theta}_{x_0 \rightarrow f}^{(1)} {\theta}_{x_1\rightarrow f}^{(1)}) \\
    LT_{w_{end}}({\theta}_{x_0\rightarrow f}^{(0)} {\theta}_{x_1 \rightarrow f}^{(1)} + {\theta}_{x_0 \rightarrow f}^{(1)} {\theta}_{x_1 \rightarrow f}^{(0)})
    \end{pmatrix}  
  \label{mu1}
\end{equation} 
where $LT_{w_{\text{end}}}(.)$ denotes the operator that only selects the monomials of a degree lower or equal to $w_{\text{end}}$.

% Figure \ref{factor_graph_u2_mu} gives an illustration of a decoding factor graph of bit $u_2$ for a polar code of length $N=8$ for  $\bp = [ 1 0 0 0 0 0 0 0]$.
% \begin{figure}[htbp]
%               \centering
%               \scalebox{0.55}{\input{Images/decoding_graph_u2_mu}}
%             \caption{Tanner Graph of $u_2$ decoding for a polar code with $N=8$ }
%             \label{factor_graph_u2_mu}
% \end{figure}

% In this example, considering the dashed sub-graph (a), we have: 
% \begin{equation}
%     \boldsymbol{\mu}_{v_0} = \begin{pmatrix}
% \min(1+1,0+0)\\
%  \min(1+0,0+1) \\
% \end{pmatrix} = \begin{pmatrix}
%  0  \\
% 1 \\
% \end{pmatrix}
% \end{equation}

Similarly, the second configuration represented in Fig. \ref{second_diagram} is where two parity nodes with parity functions $f_0$ and $f_1$ are connected to a variable node $x$. 
$T_{f_i} $ is defined as $T_{f_i} = \left\{ \bx | \quad \bx {H^T_{T_{f_i}}} = 0 \right\}$ and to each message coming from a parity node, we associate:
\begin{equation}
    \boldsymbol{\theta}_{f_i \rightarrow x} = \begin{pmatrix}
            \theta^{(0)}_{f_i \rightarrow x} \\
            \theta^{(1)}_{f_i \rightarrow x} 
        \end{pmatrix} = \begin{pmatrix}
            A_N^{w_{end}}(T_{f_i}| x = 0)(X) \\
            A_N^{w_{end}}(T_{f_i}| x = 1)(X) 
        \end{pmatrix}
\end{equation}
Given the two incoming messages $\boldsymbol{\theta}_{f_0 \rightarrow x}$ and $\boldsymbol{\theta}_{f_1 \rightarrow x}$ from the parity nodes to the variable node $x$, $\boldsymbol{\theta}_{x \rightarrow f_2}$ can be expressed as follows:
\begin{equation}
 \boldsymbol{\theta}_{x \rightarrow f_2}  = \begin{pmatrix}
        LT_{w_{end}}( {\theta}_{f_0 \rightarrow x}^{(0)} {\theta}_{f_1 \rightarrow x}^{(0)} ) \\
    LT_{w_{end}}( {\theta}_{f_0 \rightarrow x}^{(1)} {\theta}_{f_1 \rightarrow x}^{(1)} )
    \end{pmatrix}  
\label{eq_variable_1}
\end{equation}
The MWEF can be computed similarly to the RWEF by replacing the $LT_{w_{end}}(.)$ operator in Equations \eqref{mu1} and \eqref{eq_variable_1} with the $LP(.)$ operator, which only selects the monomial of lowest degree.
Finally, the initial message that is sent from a leaf node  $x_i$ is:
\begin{equation}
    \boldsymbol{\theta}_{x_i} = \begin{pmatrix}
X^{p_i \oplus 0} \\
X^{p_i \oplus 1}
\end{pmatrix}
\label{eq_initialisation}
\end{equation}

% As an example, when considering the dashed sub-graph (b) in figure \ref{factor_graph_u2_mu}, we have: 
% \begin{equation}
%     \boldsymbol{\mu}_{v_1} = \begin{pmatrix}
% 1+0  \\
% 1+0   \\
% \end{pmatrix} = \begin{pmatrix}
%  1   \\
% 1 \\
% \end{pmatrix}
% \end{equation}

\subsection{RWEF of punctured/shortened polar cosets}
\label{sec_punc_short}
 In this section, we adapt the computation of the RWEF of polar cosets in order to take the puncturing or/and the shortening effect into account. We define  rate-compatible punctured and shortened polar cosets respectively as:
\begin{equation}
 \left\{\begin{matrix}
 {\mathcal{CP}_N}(\bu_{0}^{i})=  \{ \bx_{\bar{\mathcal{P}}} | \bx \in  {\mathcal{C}_N}(\bu_{0}^{i}) \}
 \\
 {\mathcal{CS}_N}(\bu_{0}^{i})=  \{ \bx_{\bar{\mathcal{S}}} |\bx \in {\mathcal{C}_N}(\bu_{0}^{i}), \bx_{\mathcal{S}} = \bf{0}  \}
\end{matrix}\right.
\end{equation}
Where $\bar{\mathcal{V}}$ denotes the complement of the set $\mathcal{V}$.
This representation is different from ${\mathcal{C}_N}(\bu_{0}^{i})$ in the way that it takes into account the effect of puncturing or shortening. 
It is possible thus to compute $A_N^{w_{end}}( {\mathcal{CP}_N}(\bu_{0}^{i}))(X)$ and $A_N^{w_{end}}( {\mathcal{CS}_N}(\bu_{0}^{i}))(X)$ using an approach that is similar to the one used to define the LLR values of rate-compatible polar codes. \\
\subsubsection{Case of puncturing}
In the case of punctured polar codes, given $x_i$ such that $i \in \mathcal{P}$, the value of $x_p$ is erased. Therefore, it does not play any role into the determination of the different codewords weights and therefore adds no weight to the final words of the rate-compatible coset. When taking this into consideration, each leaf node $x_i, i \in \mathcal{P}$ is initialised as follows:
\begin{equation}
      \boldsymbol{\theta}_{x_i} = \begin{pmatrix}
 1 X^0 \\
1 X^0
\end{pmatrix} = \begin{pmatrix}
 1  \\
1 
\end{pmatrix}
\end{equation}
 \\
However, there is a modification that needs to be taken into consideration in the case of punctured polar codes. Actually, as a punctured polar coset ${\mathcal{CP}_N}(\bu_0^i)$ describes the affine space generated by the punctured last $N-i-1$ rows of the generator matrix, the resulting punctured matrix $\mathbf{GP}_{i+1}^{N-1}$ may not be full rank due to puncturing. This results into taking into account a word from the coset more than once. Therefore, in the case of punctured polar codes, the number of occurrences of words with weight $w$ has to be divided by $2^{N-i-1-rk(\mathbf{GP}_{i+1}^{N-1})}$, 
% \begin{equation}
%     A_N^{w_{end}}({\mathcal{C}_N}(\bu_0^i))(X) = \frac{1}{2^{N-i-1-rk(GP_{i+1}^{N-1})}}   A_N^{w_{end}}({\mathcal{C}_N}(\bu_0^i))(X)
%     \label{eq_rank_punctured}
% \end{equation}
where $rk(.)$ computes the rank of a matrix.
\begin{example}
    An illustration of a weight factor graph of bit $u_3$ is provided by Fig. \ref{factor_graph_u3_complete}. We consider the punctured polar codes with parameters $N = 8$, $K = 2$ and $P = 4$,  $\mathcal{P} = \left\{0, 2, 4, 6\right\}$, the frozen bit set $\mathcal{F} = \left\{0, 1, 2, 3,  4, 6\right\}$ and $\bp = [0, 0, 0] G_{0}^{2} = [ 0, 0, 0, 0, 0, 0, 0, 0]$. The aim is to compute the RWEFs $ A_8^2(\mathcal{CP}_8([0,0,0], u_3 = 0))$ and $ A_8^2(\mathcal{CP}_8([0,0,0], u_3 = 1))$ for $w_{end} = 2$. The RWEFs on the different nodes are computed using Equations \eqref{mu1} and \eqref{eq_variable_1}. The monomials with a power greater than $2$ are discarded.
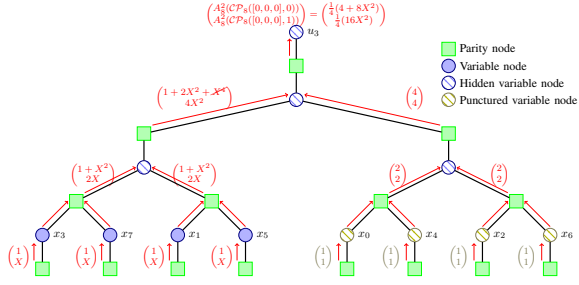
\begin{figure}[htbp]
              \centering
              \scalebox{0.45}{\pagecolor{white}

\begin{tikzpicture} 

% Style definitions
\tikzset{variable_node_style/.style={regular polygon,regular polygon sides=4,draw=green,fill=green!30!white, inner sep = 4pt}};
\tikzset{check_node_style/.style={circle,draw=blue!50!black,fill=blue!30!white, inner sep = 4pt}};
\tikzset{hidden_check_node_style/.style={circle,draw=blue!50!black,fill=blue!30!white, pattern=north west lines,pattern color=blue!50!white, inner sep = 4pt}};
\tikzset{edge_style/.style={draw=black,line width=1pt }};

\tikzset{punctured_check_node_style/.style={circle,draw=yellow!40!black,fill=yellow!30!black, pattern=north west lines,pattern color=yellow!70!black, inner sep = 4pt}};
\tikzset{edge_style/.style={draw=black,line width=1pt }};

% Variable Nodes

\node [hidden_check_node_style, label=right:{}] (c1) at (7.5,6) {};

\node [variable_node_style] (c11) at (1,3) {};
\node [variable_node_style] (c12) at (5,3) {};

\node [variable_node_style] (c21) at (10,3) {};
\node [variable_node_style] (c22) at (14,3) {};

\node [variable_node_style] (cc13) at (0,1) {};
\node [variable_node_style] (cc14) at (2,1) {};

\node [variable_node_style] (cc11) at (4,1) {};
\node [variable_node_style] (cc12) at (6,1) {};

\node [variable_node_style] (cc21) at (9,1) {};
\node [variable_node_style] (cc22) at (11,1) {};
\node [variable_node_style] (cc23) at (13,1) {};
\node [variable_node_style] (cc24) at (15,1) {};

\node [variable_node_style,label=right:{}] (f1) at (3,5) {} ;
\node [variable_node_style,label=right:{}] (f2) at (12,5) {};
\node [below] at (-0.7,1.8) {\textcolor{red}{$\text{\footnotesize $ \begin{pmatrix}
 1  \\ 
 X \\
\end{pmatrix}$}$}};

\node [below] at (1.3,1.8) {\textcolor{red}{$\text{\footnotesize $ \begin{pmatrix}
 1   \\ 
 X \\
\end{pmatrix}$}$}};

\node [below] at (3.3,1.8) {\textcolor{red}{$\text{\footnotesize $ \begin{pmatrix}
 1   \\ 
 X  \\
\end{pmatrix}$}$}};

\node [below] at (5.3,1.8) {\textcolor{red}{$\text{\footnotesize $ \begin{pmatrix}
 1   \\ 
 X  \\
\end{pmatrix}$}$}};

\node [below] at (8.3,1.8) {\textcolor{yellow!40!black}{$\text{\footnotesize $ \begin{pmatrix}
 1   \\ 
 1  \\
\end{pmatrix}$}$}};

\node [below] at (10.3,1.8) {\textcolor{yellow!40!black}{$\text{\footnotesize $ \begin{pmatrix}
 1   \\ 
 1  \\
\end{pmatrix}$}$}};

\node [below] at (12.3,1.8) {\textcolor{yellow!40!black}{$\text{\footnotesize $ \begin{pmatrix}
 1   \\ 
 1 \\
\end{pmatrix}$}$}};

\node [below] at (14.3,1.8) {\textcolor{yellow!40!black}{$\text{\footnotesize $ \begin{pmatrix}
 1   \\ 
 1  \\
\end{pmatrix}$}$}};

\node [below] at (1.5,4.25) {\textcolor{red}{$\text{\footnotesize $ \begin{pmatrix}
 1 + X^2  \\ 
 2X  \\
\end{pmatrix}$}$}};

\node [below] at (4.5,4.25) {\textcolor{red}{$\text{\footnotesize $ \begin{pmatrix}
 1 + X^2   \\ 
 2X  \\
\end{pmatrix}$}$}};

\node [below] at (10.5,4.25) {\textcolor{red}{$\text{\footnotesize $ \begin{pmatrix}
 2   \\ 
 2  \\
\end{pmatrix}$}$}};

\node [below] at (13.5,4.25) {\textcolor{red}{$\text{\footnotesize $ \begin{pmatrix}
 2   \\ 
 2  \\
\end{pmatrix}$}$}};

\node [below] at (4.5,6.5) {\textcolor{red}{$\text{\footnotesize $ \begin{pmatrix}
1 +2 X^2 + \cancel{X^4 } \\ 
 4X^2 \\
\end{pmatrix}$}$}};

\node [below] at (11,6.5) {\textcolor{red}{$\text{\footnotesize $ \begin{pmatrix}
 4    \\ 
 4 \\
\end{pmatrix}$}$}};
\node [below] at (7.5,9) {\textcolor{red}{$\text{\footnotesize $ \begin{pmatrix}
 A_8^2(\mathcal{CP}_8([0,0,0], 0))   \\ 
 A_8^2(\mathcal{CP}_8([0,0,0], 1))  \\
\end{pmatrix} = \begin{pmatrix}
  \frac{1}{4}(4 + 8 X^2)  \\ 
 \frac{1}{4} (16X^2)  \\
\end{pmatrix}$}$}};
% Check Nodes
\node [hidden_check_node_style,label=right:{}] (x1) at (3,4) {} ;
\node [hidden_check_node_style,label=right:{}] (x2) at (12,4) {};
\node [variable_node_style,label=right:{}] (x3) at (7.5,7) {};
\node [hidden_check_node_style,label=right:{$u_3$}] (xf) at (7.5,8) {};
\node [check_node_style,label=right:{$x_1$}] (x11) at (4,2) {};
\node [check_node_style,label=right:{$x_5$}] (x12) at (6,2) {};
\node [check_node_style,label=right:{$x_3$}] (x13) at (0,2) {};
\node [check_node_style,label=right:{$x_7$}] (x14) at (2,2) {};

\node [punctured_check_node_style,label=right:{$x_0$}] (x21) at (9,2) {};
\node [punctured_check_node_style,label=right:{$x_4$}] (x22) at (11,2) {};
\node [punctured_check_node_style,label=right:{$x_2$}] (x23) at (13,2) {};
\node [punctured_check_node_style,label=right:{$x_6$}] (x24) at (15,2) {};

% Edges
% \path [edge_style] (x1) -- (c1);
% \path [edge_style] (x2) -- (c1);
\path [edge_style] (x3) -- (c1);

\path [edge_style] (x1) -- (c11);
\path [edge_style] (x1) -- (c12);

\path [edge_style] (x2) -- (c21);
\path [edge_style] (x2) -- (c22);

\path [edge_style] (c12) -- (x11);
\path [edge_style] (c12) -- (x12);
\path [edge_style] (c11) -- (x13);
\path [edge_style] (c11) -- (x14);

\path [edge_style] (c21) -- (x21);
\path [edge_style] (c21) -- (x22);
\path [edge_style] (c22) -- (x23);
\path [edge_style] (c22) -- (x24);

\path [edge_style] (cc11) -- (x11);
\path [edge_style] (cc12) -- (x12);
\path [edge_style] (cc13) -- (x13);
\path [edge_style] (cc14) -- (x14);

\path [edge_style] (cc21) -- (x21);
\path [edge_style] (cc22) -- (x22);
\path [edge_style] (cc23) -- (x23);
\path [edge_style] (cc24) -- (x24);

\path [edge_style] (x2) -- (f2);
\path [edge_style] (x1) -- (f1);

 \path [edge_style] (f1) -- (c1);
 \path [edge_style] (f2) -- (c1);
  \path [edge_style] (xf) -- (x3);

% \draw [dashed] (-0.5,3.5) -- (2.5,3.5) -- (2.5,1.5)--(-0.5,1.5) -- cycle  ; 
% \node [below] at (2.75,3.75) {$(a)$};

% \draw [dashed] (9.5,4.5) -- (14.5,4.5) -- (14.5,2.5)--(9.5,2.5) -- cycle  ; 
% \node [below] at (10.25,5) {$(b)$};

%% arrows 
\draw  [->] [color=red](0,2.25) -- (0.75,3);
\draw  [->] [color=red](2,2.25) -- (1.25,3);

\draw  [->] [color=red](4,2.25) -- (4.75,3);
\draw  [->] [color=red](6,2.25) -- (5.25,3);

\draw  [->] [color=red](9,2.25) -- (9.75,3);
\draw  [->] [color=red](11,2.25) -- (10.25,3);

\draw  [->] [color=red](13,2.25) -- (13.75,3);
\draw  [->] [color=red](15,2.25) -- (14.25,3);

\draw  [->] [color=red](1.25,3.25) -- (2.75,4);
\draw  [->] [color=red](4.75,3.25) -- (3.25,4);

\draw  [->] [color=red](10.25,3.25) -- (11.75,4);
\draw  [->] [color=red](13.75,3.25) -- (12.25,4);

\draw  [->] [color=red](3.25,5.25) -- (7.25,6.15);
\draw  [->] [color=red](11.75,5.25) -- (7.75,6.15);
\draw  [->] [color=red](-0.25,1.25) -- (-0.25,1.75);
\draw  [->] [color=red](1.75,1.25) -- (1.75,1.75);
\draw  [->] [color=red](3.75,1.25) -- (3.75,1.75);
\draw  [->] [color=red](5.75,1.25) -- (5.75,1.75);
\draw  [->] [color=red](8.75,1.25) -- (8.75,1.75);
\draw  [->] [color=red](10.75,1.25) -- (10.75,1.75);
\draw  [->] [color=red](12.75,1.25) -- (12.75,1.75);
\draw  [->] [color=red](14.75,1.25) -- (14.75,1.75);

\draw  [->] [color=red](7.3,7.3) -- (7.3,7.8);

%% Legend 
\node [variable_node_style, label=right:{Parity node}] (x11) at (12,7.5) {};
\node [check_node_style, label=right:{Variable node}] (x11) at (12,7) {};
\node [hidden_check_node_style, label=right:{Hidden variable node}] (x11) at (12,6.5) {};

\node [punctured_check_node_style, label=right:{Punctured variable node}] (x11) at (12,6) {};

\end{tikzpicture}
 }
            \caption{Tanner Graph of $u_3$ decoding for a punctured polar code with $N = 8$ and $P = 4$}
            \label{factor_graph_u3_complete}
\end{figure}

It has also to be noted that the cosets ${\mathcal{CP}_8}([0, 0, 0], 0)$ and ${\mathcal{CP}_8}([0, 0, 0], 1)$ describe the space generated by the  punctured rows of the matrix highlighted in green in Fig. \ref{fig_punctured_matrix}. 
\begin{figure}[h]
  \centering
  \includegraphics[scale=0.7]{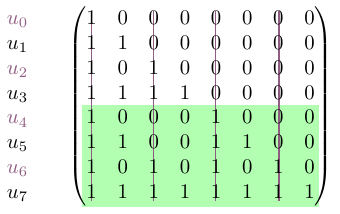}
  \caption{Transformation matrix for $N =8$ and $P = 4$}
            \label{fig_punctured_matrix}
\end{figure}
We can see from Fig. \ref{fig_punctured_matrix} that due to the presence of two zero rows (the fifth and the seventh rows), the rank of the matrix  $\mathbf{GP}_{4}^{7}$ is equal to 2 instead of 4 when taking the puncturing into account. This means that $A_8^2(\mathcal{CP}_8([0,0,0], 0)) = \frac{1}{2^{8 - 3-1-2}}(4 +8 X^2) = 1 + 2 X^2$ and ${\mathcal{CP}_8}([0, 0, 0],1) = \frac{1}{2^{8 - 3-1-2}}.16X^2 = 4X^2$.
\end{example}
\subsubsection{Case of shortening}
In the case of shortened polar codes, the shortening pattern is defined to guarantee that $x_i = 0, i \in \mathcal{S}$. This means that there are no codewords configurations for which $x_i = 1$. This leads to the following initialization for every leaf node $x_i$ such that $i \in \mathcal{S}$:
\begin{equation}
      \boldsymbol{\theta}_{x_i} = \begin{pmatrix}
 1 \\
0
\end{pmatrix}
\end{equation}
% As the second row of $\boldsymbol{\theta}_{x_i}$ defines the weight of the relative configuration to $x_i$ when $x_i$ is equal to 1, setting it to 0 prevents it from appearing in any configuration with a given weight. 
The message passing rules remain unchanged.
% Fig. \ref{factor_graph_u2_mu}. gives an illustration of a decoding factor graph of bit $u_3$. We consider configurations for a shortened polar code. In this configuration, $N = 8$, $K = 4$ and $S = 2$. The shortening pattern $\mathcal{S} = \left\{3, 7\right\}$ and the frozen bit set $\mathcal{F} = \left\{0, 1, 3, 7\right\}$. The evaluated coset is ${\mathcal{CS}^{(2)}_8}([0, 1, 0])$, thus $\bp = [0, 1, 0] G_{0}^{2} = [ 1 1 0 0 0 0 0 0]$.
% \begin{figure}[htbp]
%               \centering
%               \scalebox{0.5}{\input{Images/decoding_graph_u3_mu}}
%             \caption{Tanner Graph of $u_3$ decoding for a shortened polar code with $N = 8$ and $S = 2$}
%             \label{factor_graph_u2_mu}
% \end{figure}
% In the case of the shortened codeword bits $x_3$ and $x_7$, they are initialised to $\begin{pmatrix}
%  0 \\
% \infty
% \end{pmatrix}$, the rest are initialized according to \eqref{eq_initialisation}. The message passing is based on Equations \eqref{mu1} and \eqref{eq_variable_1}
% We can see from Fig. \ref{factor_graph_u2_mu}. that $\boldsymbol{\mu}_{u_3}  = \begin{pmatrix}
% 2   \\
% \infty
% \end{pmatrix}$. This means that $w^*({\mathcal{CS}^{(2)}_8}([0, 1, 0],0)) = 2$ for $u_3 = 0$. On the other hand, $w^*({\mathcal{CS}^{(2)}_8}([0, 1, 0],1)) = \infty$, which reflects the fact that the initialization takes into account the fact that there are no configurations for the shortened polar code for which $u_3 = 1$ (since $u_3$ is forced to $0$ due to the shortening pattern).

\section{Enumeration of low-weight codewords for rate-compatible pre-transformed polar codes}
\label{relaxed_polar}

%After showing in Section \ref{sec_min_weight} that we are able to compute the minimum weight any polar coset ${\mathcal{CV}^{(i)}_N}(\bu_{0}^{i-1},u_i)$, accounting for the impact of puncturing and shortening. 
In this section, we leverage the findings from Section \ref{sec_min_weight} to enumerate all the codewords of weight less or equal to a fixed threshold of punctured and shortened pure and pre-transformed polar codes.
\subsection{Rate-compatible polar codes as union of rate-compatible cosets}
% It has been shown in  in \cite{b6} that any polar code $\mathcal{C}$ can be formulated as the disjoint union of the following cosets:
% \begin{equation}
%  \mathcal{C} = \biggcup_{\bu_{0}^{s} \in L_s} \mathcal{C}^{(s)}_N(\bu_{0}^{s-1},u_{s} = 0) 
% \label{eq:first_union}
% \end{equation}
% Where $s$ denotes the index of the last frozen bit and the set $Ls = \left\{
% \bu_0^{s-1} \in {\{0,1\}}^s | u_i =0, \forall i \in \mathcal{F}
% \right\}$.\\
% We will show, that in the case of punctured or shortened polar codes, a disjoint union of the rate-compatible cosets describes the whole rate-compatible polar code. 
\begin{example}
    Let us consider the polar code shortened from the parent code (16,7) with $\mathcal{S} = \left\{7, 15\right\}$. Fig. \ref{fig_polar_transformation} shows the transformation matrix $\bG_{16}$. 
    \begin{figure}[htbp]
              \centering
              \scalebox{0.5}{\pagecolor{white}

\begin{tikzpicture} 
\myRectangle{0.45}{-1.78}{8}{0.33}{green!30!white}{green!30!white}
\myRectangle{0.45}{1.55}{8}{0.33}{green!30!white}{green!30!white}

\myRectangle{4.15}{-1.7}{0.33}{6.6}{green!30!white}{green!30!white}
\myRectangle{8.35}{-1.7}{0.33}{6.6}{green!30!white}{green!30!white}

\node [below] at (-1,5) {$
\begin{matrix}
u_0  \\
u_1  \\
u_2  \\
u_3  \\
u_4  \\
u_5  \\
\textcolor{red}{u_6}\\
\textcolor{green!70!black}{u_7} \\
u_8\\
\textcolor{red}{u_9} \\
\textcolor{blue}{u_{10}} \\
\textcolor{blue}{u_{11}}  \\
\textcolor{blue}{u_{12}}  \\
\textcolor{blue}{u_{13}} \\
\textcolor{blue}{u_{14}}\\
\textcolor{green!70!black}{u_{15}} 
\end{matrix}
$};

\node [below] at (4.6,5) {$
\begin{pmatrix}
1 & 0 & 0 & 0 & 0 & 0 & 0 & 0 & 0 & 0 & 0 & 0 & 0 & 0 & 0 & 0 \\
1 & 1 & 0 & 0 & 0 & 0 & 0 & 0 & 0 & 0 & 0 & 0 & 0 & 0 & 0 & 0 \\
1 & 0 & 1 & 0 & 0 & 0 & 0 & 0 & 0 & 0 & 0 & 0 & 0 & 0 & 0 & 0 \\
1 & 1 & 1 & 1 & 0 & 0 & 0 & 0 & 0 & 0 & 0 & 0 & 0 & 0 & 0 & 0 \\
1 & 0 & 0 & 0 & 1 & 0 & 0 & 0 & 0 & 0 & 0 & 0 & 0 & 0 & 0 & 0 \\
1 & 1 & 0 & 0 & 1 & 1 & 0 & 0 & 0 & 0 & 0 & 0 & 0 & 0 & 0 & 0 \\
1 & 0 & 1 & 0 & 1 & 0 & 1 & 0 & 0 & 0 & 0 & 0 & 0 & 0 & 0 & 0 \\
1 & 1 & 1 & 1 & 1 & 1 & 1 & 1 & 0 & 0 & 0 & 0 & 0 & 0 & 0 & 0 \\
1 & 0 & 0 & 0 & 0 & 0 & 0 & 0 & 1 & 0 & 0 & 0 & 0 & 0 & 0 & 0 \\
1 & 1 & 0 & 0 & 0 & 0 & 0 & 0 & 1 & 1 & 0 & 0 & 0 & 0 & 0 & 0 \\
1 & 0 & 1 & 0 & 0 & 0 & 0 & 0 & 1 & 0 & 1 & 0 & 0 & 0 & 0 & 0 \\
1 & 1 & 1 & 1 & 0 & 0 & 0 & 0 & 1 & 1 & 1 & 1 & 0 & 0 & 0 & 0 \\
1 & 0 & 0 & 0 & 1 & 0 & 0 & 0 & 1 & 0 & 0 & 0 & 1 & 0 & 0 & 0 \\
1 & 1 & 0 & 0 & 1 & 1 & 0 & 0 & 1 & 1 & 0 & 0 & 1 & 1 & 0 & 0 \\
1 & 0 & 1 & 0 & 1 & 0 & 1 & 0 & 1 & 0 & 1 & 0 & 1 & 0 & 1 & 0 \\
1 & 1 & 1 & 1 & 1 & 1 & 1 & 1 & 1 & 1 & 1 & 1 & 1 & 1 & 1 & 1 
\end{pmatrix}
$};

\end{tikzpicture}
 }
            \caption{Polar transformation $\bG_{16}$ }
            \label{fig_polar_transformation}
\end{figure}
    In this figure, the frozen bits of indexes $\mathcal{F}= \left\{0, 1, 2, 3, 4, 5, 8\right\}$, are represented in black, the shortened bits in green and the remaining information bits are represented in red for the ones before the last frozen bit and in blue for the ones after the last frozen bit.\\
    Let us consider the shortened polar coset ${\mathcal{CS}_{16}}(\bu_{0}^{8})$. For any $\bu_0^8$ such that $u_i = 0 \quad \forall i \in \mathcal{F}$, and $u_i = \left\{0, 1\right\} \quad \forall i \in \mathcal{I}$. The shortened polar coset ${\mathcal{CS}_{16}}(\bu_{0}^{8})$  forms a subset of the shortened polar code $\mathcal{CS}$ as by definition of ${\mathcal{CS}_{16}}(\bu_{0}^{8})$, the shortened bits $u_7$ and $u_{15}$ are constrained to be zero. Therefore, the total shortened polar code $\mathcal{C}_{\mathcal{S}}$ can be described as: 
    \begin{equation}
        \mathcal{C}_{\mathcal{S}} = \biggcup_{u_0^8 \in L_8} {\mathcal{CS}_{16}}(\bu_{0}^{8})
    \end{equation}
    
    Where 
    $L_8 = \left\{\bu_0^8 \ in \left\{0,1\right\}^8| u_i = 0 \quad \forall i \in \mathcal{F} \right\}$. 
\end{example}
More generally, any  shortened polar code $\mathcal{C}_{\mathcal{S}}$ can be expressed as: 
\begin{equation}
        \mathcal{C}_{\mathcal{S}} = \biggcup_{\bu_0^s \in L_s} {\mathcal{CS}_{N}}(\bu_{0}^{s})
    \end{equation}
where $s = \text{max}(F)$  and $L_s =  \left\{\bu_0^s \ in \left\{0,1\right\}^s| u_i = 0 \quad \forall i \in \mathcal{F} \right\}$. 
The RWEF $A_N^{w_{end}}(\mathcal{C}_{\mathcal{S}})(X)$ of the shortened polar code is therefore obtained as: 
\begin{equation}
    A_N^{w_{end}}(\mathcal{C}_{\mathcal{S}})(X) = \sum\limits_{\bu_0^s \in L_s} A_N^{w_{end}}({\mathcal{CS}_{N}}(\bu_{0}^{s}))
\end{equation}
    The same reasoning remains valid in the case of a  punctured polar code $\mathcal{C}_{\mathcal{P}}$. \\
    Exploring all the possible cosets may become prohibitive even for moderate code sizes. We propose in the next section an algorithm that only explores relevant cosets. \\
\subsection{Reduced spectrum of rate-compatible polar codes}
% The aim is to take advantage of the computation of the minimum weight of the rate-compatible polar cosets in order to eliminate the irrelevant cosets for the computation of the partial spectrum and to only keep the codewords having a weight less or equal to $w_{end}$. \\
We propose an algorithm that deterministically computes the partial spectrum up to a specified weight $w_{end}$ while restricting the exploration to only the relevant cosets during the process.

We are able to compute the minimum weight  $w^*$  of any rate-compatible polar coset ${\mathcal{CV}_N}(\bu_{0}^{i})$, where $\mathcal{V}$ represents $\mathcal{S}$ when dealing with shortening and $\mathcal{P}$ when handling puncturing. This is achieved by calculating the MWEF of the coset. Therefore, it is possible to propose an enumeration structure that has the advantage of pruning cosets with a constraint of their minimal weight. The fixed value of the last explored weight $w_{end}$ is used as a threshold to eliminate irrelevant prefixes. This operates as follows:
\begin{itemize}
    \item For each $i \in [0,N-1] $, all the prefixes $\bu_{0}^{i}$ where $ u_i = \bg(\bv_0^i)$ that remained in the list at an exploration stage $i$ are listed.
    \item For each of the aforementioned prefixes,  $w^*({\mathcal{CV}_N}(\bu_{0}^{i}))$ is computed.
\item The cosets with $w^*({\mathcal{CV}_N}(\bu_{0}^{i}))>w_{end}$ are discarded. Those cosets are irrelevant to the computation the partial weight spectrum with the threshold on weight $w_{end}$ as:
\begin{equation}
w^*({\mathcal{CV}_N}(\bu_{0}^{i},u_{i+1})) \geqslant  w^*({\mathcal{CV}_N}(\bu_{0}^{i-1},u_{i}) )
\end{equation}
% This means that $\forall \quad {\mathcal{CV}_N}(\bu_{0}^{i},u_{i+1}) $ with $w^*({\mathcal{CV}_N}(\bu_{0}^{i},u_{i+1}))< w_{end}$, $\nexists j \in [\![i+1;N-1]\!]$ such that $w^*({\mathcal{CV}_N}(\bu_{0}^{j-1},u_{j})) = w_{end}$. 
In other words, if a coset has a minimum weight $w > w_{end}$, then no codeword within that coset can have a weight lower than or equal to $w_{end}$. \\
\item When $i=s$, the partial weight spectrum is obtained as the sum of the RWEFs of the cosets remaining in the list.
    
\end{itemize}
Algorithm \ref{algo} gives the details of the proposed algorithm for punctured or shortened PAC codes. It is important to note that it can also be applied to polar codes with DFB as this only affects the way the pre-transformation is realised.  Algorithm \ref{algo} consists of $s-1$ loop iterations where the minimum weight of $C_i$ cosets is evaluated at each enumeration stage and one iteration where the RWEF of $C_s$ cosets is evaluated. The computational complexity of the proposed method is driven by the total number of evaluated cosets $ n_c = \sum_{i=0}^{s}C_i$.
\begin{algorithm}[!h]
\SetAlgoLined
\DontPrintSemicolon
\KwIn{ $N, K, \mathcal{F}, \bg, \mathcal{V}, w_{end}$}
\KwOut{ Reduced spectrum up to $w_{end}$}
$L \leftarrow 1$\;
$\mathcal{L} \leftarrow \left\{0\right\}$ \tcc{List to store prefixes}\;  
 \For{$i \in [\![0;s]\!]$}{
    \uIf{$i \in \mathcal{F}$}{
        \For{$l \in [\![1;L]\!]$}
        {
        $v_i[l] \leftarrow 0$\;
        \uIf{$i \in \mathcal{S}$}{
         $u_i[l] \leftarrow 0$ \tcc{Equation \eqref{eq_shortened_pac}}\;
        }
        \Else{
        $u_i[l] \leftarrow \bg(\bv_0^i[l])$\;
        }
        Compute $w^{*}$   of ${\mathcal{CV}_N}(\bu_{0}^{i-1}[l],u_i[l])$\;
 Discard the prefixes for which $w^{*} > w_{end}$\;
 $L \leftarrow |\mathcal{L}|$\;
        } 
    }
    \Else{
        $\mathcal{L} \leftarrow \mathcal{L} \bigcup \mathcal{L'}$ \tcc{$\mathcal{L'}$ is a copy of $\mathcal{L'}$}\;
        \For{$l \in [\![1;L]\!]$  }{
            $[v_i[l], v_i[l']] \leftarrow [0, 1]$\;
            $u_i[l] \leftarrow \bg(\bv_0^i[l])$ $u_i[l'] \leftarrow \bg(\bv_0^i[l'])$\;
            Compute $w^{*}$   of ${\mathcal{CV}_N}(\bu_{0}^{i-1}[l],u_i[l])$ and ${\mathcal{CV}_N}(\bu_{0}^{i-1}[l],u_i[l'])$ \;
            Discard the cosets for which $w^{*} > w_{end}$\;
            $L \leftarrow |\mathcal{L}|$\;
            
        }
        }

            \uIf{$i = s$}{
            Compute the RWEF of the remaining paths in the list \;
            Compute the RWEF of the overall code $A_N^{w_{end}}(\mathcal{C}_{\mathcal{V}})$ 
            
            }
 }

    {
    Return $A_N^{w_{end}}(\mathcal{C}_{\mathcal{V}})$\;
}

 \caption{Reduced spectrum of punctured/shortened PAC codes }
 \label{algo}
 
\end{algorithm}
\begin{table}[]
\scriptsize
\centering
\caption{Partial weight distribution of punctured and shortening polar and PAC codes}
\label{tab3}
\setlength\extrarowheight{2pt}
\begin{tabular}{|l|l|l|}
\hline
$(N,K)$                                                                        & Type  & $(w, A_w)$                                                                                                                                                                                                                                               \\ \hline
\multirow{2}{*}{$(80, 20)$} & Polar & \begin{tabular}[c]{@{}l@{}}$(8, 30), (16, 173), (20, 256) (24, 8040)$\\  $(28, 7424)$\end{tabular}                                                                                                                                                           \\ \cline{2-3} 
                         & PAC   & \begin{tabular}[c]{@{}l@{}}$(8, 2), (12, 8), (14, 8), (16, 109), (18, 56)$\\  $(20, 920), (22, 504), (24, 2456)$\\ $(26, 5352), (28, 11528), (30, 11304)$\\ $(32, 34194)$\end{tabular}                                                                           \\ \hline
\multirow{2}{*}{$(160,40)$}                                                    & Polar & \begin{tabular}[c]{@{}l@{}}$(8,12), (16,382), (24, 2220), (32, 55533)$\\  $(40, 663536)$\end{tabular}                                                                                                                                                        \\ \cline{2-3} 
                         & PAC   & \begin{tabular}[c]{@{}l@{}}$(8, 8), (16, 230), (18, 64), (20, 64)$\\ $(22, 64), (24, 2120), (26, 960)$\\ $(28, 1216), (30, 1472),(32, 16557)$\\ $(34, 16448), (36, 19264)$\\ $(38, 22080), (40, 286240)$\end{tabular}                                              \\ \hline
\multirow{2}{*}{$(320, 80)$}                                                   & Polar & \begin{tabular}[c]{@{}l@{}}$(16, 476), (24, 11584), (28, 12288)$\\ $(32, 117598), (36, 12288)$\\ $(40, 678208), (44, 589824)$\\ $(48, 15764476)$\end{tabular}                                                                                                    \\ \cline{2-3} 
                         & PAC   & \begin{tabular}[c]{@{}l@{}}$(16, 76), (18, 16), (20, 32), (22, 16)$\\  ($24, 208), (26, 112), (28, 352), (30, 112)$\\  $(32, 9694), (34, 2928), (36, 12512)$\\  $(38, 8176), (40, 160848), (42, 52496)$\\ $(44, 224544), (46, 192784), (48, 3669484)$\end{tabular} \\ \hline
\multirow{2}{*}{$(640, 160)$}                                                  & Polar & \begin{tabular}[c]{@{}l@{}}$(16, 344), (24, 2688), (32, 117004)$\\ $(40, 3741824)$\end{tabular}                                                                                                                                                              \\ \cline{2-3} 
                         & PAC   & \begin{tabular}[c]{@{}l@{}}$(16, 20), (24, 24), (32, 11652), (34, 64)$\\ $(36, 704), (38, 64), (40, 47032)$\end{tabular}                                                                                                                                     \\ \hline
\multirow{2}{*}{$(80, 40)$}                                                    & Polar & \begin{tabular}[c]{@{}l@{}}$\textcolor{red}{(8, 1078)}, (12, 32128), (14, 45056)$\\  $(16, 971821), (18, 2191360)$\\  $(20, 35615872)$\end{tabular}                                                                                                                             \\ \cline{2-3} 
                         & PAC   & \begin{tabular}[c]{@{}l@{}}$\textcolor{red}{(8, 582)}, (10, 608),  (12, 14848)$\\  $(14, 46624), (16, 446125), (18, 1810400)$\\ $ (20, 11718144)$\end{tabular}                                                                                                                  \\ \hline
\multirow{2}{*}{$(160, 80)$}                                                   & Polar & \begin{tabular}[c]{@{}l@{}}$\textcolor{red}{(8, 508)}, (12, 2496), (16, 320030)$\\  $(20, 9821632)$\end{tabular}                                                                                                                                                              \\ \cline{2-3} 
                         & PAC   & \begin{tabular}[c]{@{}l@{}}$\textcolor{red}{(8, 300)}, (12,2112), (16, 92862)$\\  $(18, 15616), (20, 2453568)$\end{tabular}                                                                                                                                                   \\ \hline
\multirow{2}{*}{$(320, 160)$}                                                  & Polar & $\textcolor{red}{(8,120)}, (16, 183116), (20, 731136) $                                                                                                                                                                                                                     \\ \cline{2-3} 
                         & PAC   & \begin{tabular}[c]{@{}l@{}}$\textcolor{red}{(8,120)}, (16, 74540), (18, 6080)$\\  $(20, 568832)$\end{tabular}                                                                                                                                                                 \\ \hline
\multirow{2}{*}{$(640, 320)$}                                                  & Polar & $(16, 69496), (24, 27166592)    $                                                                                                                                                                                                                          \\ \cline{2-3} 
                         & PAC   & \begin{tabular}[c]{@{}l@{}}$(16, 43544), (18, 736), (20, 25408)$\\ $(22, 6208), (24, 16013568)$\end{tabular}                                                                                                                                                         \\ \hline
\end{tabular}
\end{table}

\begin{table}
\scriptsize
\centering
    \caption{Partial weight distribution of $(200,100)$ randomly punctured / shortened polar codes}
\label{tab2}
\begin{tblr}{
  cell{2}{1} = {r=13}{},
  vlines,
  hline{1-2,15-17} = {-}{},
  hline{3-15} = {2-3}{},
}
$(N,K) $                                     & {$(200,100)$\\Random Shortening} & {$(200,100)$\\ Random puncturing} \\
$(w, A_w)$                                   & $(8,194)$                        & $(7,12)$                          \\
                                             & $(12,456)$                       & $(8,35)$                          \\
                                             & $(16,67867)$                     & $(9,115)$                         \\
                                             & $(20,1319413)$                   & $(10,332)$                        \\
                                             & $(22,696208)$                    & $(11,710)$                        \\
                                             &                                  & $(12,1349)$                       \\
                                             &                                  & $(13,2934)$                       \\
                                             &                                  & $(14,6737)$                       \\
                                             &                                  & $(15,16490)$                      \\
                                             &                                  & $(16,41033)$                      \\
                                             &                                  & $(17,98835)$                      \\
                                             &                                  & $(18,235252)$                     \\
                                             &                                  & $(19,561588)$                     \\
TC \cite{powerful} & $56257 \times 10^9$              & $128664 \times 10^9$              \\
Our TC                                 & $ 78 \times 10^9$                & $ 48 \times 10^9$                 
\end{tblr}
\end{table}
\section{Experimental results}
This section summarizes the experimental results obtained on the partial weight distribution for a wide range of pure and pre-transformed rate-compatible polar codes. For each code, we compute the exact number $A_w$ of codewords of weight $w$ for all $ w \leq w_{end}$.\\
% Table \ref{tab3} lists the minimum distance $d^*$, the associated number of occurrences $A^*$ as well as the total number of explored cosets $n_d$ of shortened pure and pre-transformed polar codes. In each case, the rates $R = 0.5$ and $R = 0.75$ are considered. In the case of PAC codes, the polynomial $\bg =  [1, 0 ,1, 1, 0, 1, 1]$ is used and the case of shortened polar codes with dynamic frozen bits, the dynamic frozen bits are determined as a random combination of previous information bits. The frozen bit sets referred to as 5G are the ones specified in the 5G standard \cite{b12} and the ones referred to as 5G-RM are the ones introduced in \cite{b244}. \\
Table \ref{tab3} summarizes the partial weight spectrum of punctured shortened polar and PAC codes for $N = \left\{128, 256, 512, 1024\right\}$, $P = \left\{48, 96, 192, 384\right\}$ and $S = \left\{48, 96, 192, 384\right\}$ respectively. We apply puncturing for the codes with rate $R = 0.25$ and shortening for codes with rate $R = 0.5$. This choice aligns with the 5G standardization, where shortening is  used for high rates and puncturing for low rates \cite{b12}. In the case of PAC codes, the polynomial $\bg =  [1, 0 ,1, 1, 0, 1, 1]$ is chosen. The frozen bit sets are the ones specified in the 5G standard \cite{b12}. The puncturing and shortening patterns are the ones defined with the bit-reversal permutation \cite{b27}.\\
The results for the number of codewords with minimum weight of shortened polar and PAC codes (results highlighted in red)  were corroborated with results in \cite{b244}. To the best of the authors' knowledge, the full results for the partial weight spectrum of PAC codes have not been reported in the literature.  \\
% Note that the proposed Algorithm \ref{algo} can be used more specifically to determine the minimum distance and associated number of occurrences of punctured and shortened polar codes when setting $w_{end}$ to the value of the minimum distance. In the case of rate-compatible polar codes, the minimum distance can be easily found using the results in \cite{b4}. In the case of pre-transformed rate-compatible polar codes, however, there is no direct way to compute the minimum distance. As the minimum distance of a pre-transformed polar code is greater or equal to the minimum distance of a pure polar code \cite{b30}, in the case of pre-transformed rate-compatible polar codes, a greedy method is used to determine $d^*$. We first set $w_{end}$ to the minimum distance of rate-compatible polar code. If no codewords can be enumerated at the last enumeration stage, $w_{end}$ is incremented by $2$ until finding codewords whose minimum distance equals $w_{end}$.
% As shown in table \ref{tab3}, PAC codes have fewer low-weight codewords. \\
Table \ref{tab2} provides a computational complexity comparison of the proposed algorithm to the one introduced in \cite{powerful}. To this end, the punctured and shortened patterns were defined randomly to accommodate the results of \cite{powerful}. The results are shown for a $(200,100)$ polar code with the same rate-profiling. We compare for both method the Time Complexity (TC) defined as the number of arithmetic operations for both methods. \\
% We proved in \cite{b23} that the computation of the minimum weight of a coset has a worst-case time complexity equal to $6(N-1)$. Algorithm \ref{algo} has therefore a worst case time complexity equal to $6n_c (N-1)$, whereas the algorithm proposed in \cite{powerful} has a time complexity dominated by the term $\Sigma {\mathcal{N}}_{Subcode}^{(\mathcal{X})}$.
Table \ref{tab2} shows that the number of codewords with a specific weight computed via Algorithm \ref{algo} is in the same range of the results computed in \cite{powerful}. Note that since the puncturing and shortening are done randomly, we cannot reproduce the exact same results. The computational complexity of Algorithm \ref{algo} is lower by several orders of magnitude. For instance , it is indicated in \cite{powerful} that the overall running time for a C++ implementation on a computer with 6 cores i7 and a 3.2GHz processor in the case of a randomly shortened $(200, 100)$ polar code is approximately 28 hours. In contrast, our MATLAB implementation on a computer with 2 cores i5 and a 3.1GHz processor achieves a running time of less than 5 minutes.

% Figure \ref{courbes_shortened} represents the Frame Error rate (FER) for shortened codes with $N = 256$ and $S = 96$ and for rates $R = 0.5$ and $R = 0.75$ under SCL decoding with $L = 32$. The result is compared to the upper bound for the FER of linear codes under ML decoding expressed as: 
% \begin{equation}
%     P_e \approx  A^* \mathcal{Q}\left(2d^* R \frac{E_b}{N_0}\right)
%     \label{eqep}
% \end{equation}
% Where $\mathcal{Q}(u)$ is defined as $\mathcal{Q}(u) = \frac{1}{\sqrt{2 \pi}} \int_{x}^{\infty}e^{-\frac{u^2}{2}} {du}$.
% \begin{figure}[htp]
%   \centering
%       \newlength\fheight
%   \newlength\fwidth
%   \setlength\fheight{7.5cm}
%   \setlength\fwidth{7.5cm}
%   \input{Images/courbe_shortened}
%   \caption{Performance evaluation for shortened polar and PAC codes }
%   \label{courbes_shortened}
% \end{figure}
% Figure shows that at high SNR ratios, the curves superimpose. 
\section{Conclusion}
A low-complexity algorithm is detailed to compute the partial weight spectrum of punctured and shortened pure and pre-transformed polar codes. The proposed approach takes advantage of the computation of cosets minimum weights to explore only the relevant cosets defining a polar code. It has been shown to have a significantly lower complexity when compared to state-of-the-art algorithms. 
Besides, the computation is feasible regardless of the frozen bit set or the punctured/shortened pattern.

\bibliographystyle{IEEEtran}
\bibliography{biblio}

\begin{thebibliography}{10}
\providecommand{\url}[1]{#1}
\csname url@samestyle\endcsname
\providecommand{\newblock}{\relax}
\providecommand{\bibinfo}[2]{#2}
\providecommand{\BIBentrySTDinterwordspacing}{\spaceskip=0pt\relax}
\providecommand{\BIBentryALTinterwordstretchfactor}{4}
\providecommand{\BIBentryALTinterwordspacing}{\spaceskip=\fontdimen2\font plus
\BIBentryALTinterwordstretchfactor\fontdimen3\font minus \fontdimen4\font\relax}
\providecommand{\BIBforeignlanguage}[2]{{%
\expandafter\ifx\csname l@#1\endcsname\relax
\typeout{** WARNING: IEEEtran.bst: No hyphenation pattern has been}%
\typeout{** loaded for the language `#1'. Using the pattern for}%
\typeout{** the default language instead.}%
\else
\language=\csname l@#1\endcsname
\fi
#2}}
\providecommand{\BIBdecl}{\relax}
\BIBdecl

\bibitem{b1}
E.~Arikan, ``Channel polarization: A method for constructing capacity-achieving codes for symmetric binary-input memoryless channels,'' \emph{IEEE Trans. on Inf. Theory}, vol.~55, no.~7, pp. 3051--3073, 2009.

\bibitem{b3}
B.~Li, H.~Shen, and D.~Tse, ``An adaptive successive cancellation list decoder for polar codes with cyclic redundancy check,'' \emph{IEEE Communications Letters}, vol.~16, no.~12, pp. 2044--2047, 2012.

\bibitem{b2}
I.~Tal and A.~Vardy, ``List decoding of polar codes,'' in \emph{2021 IEEE International Symposium on Information Theory (ISIT)}, 2011, pp. 1--5.

\bibitem{b14}
P.~Trifonov and V.~Miloslavskaya, ``Polar codes with dynamic frozen symbols and their decoding by directed search,'' in \emph{2013 IEEE Information Theory Workshop (ITW)}, 2013, pp. 1--5.

\bibitem{b13}
E.~Ar{\i}kan, ``From sequential decoding to channel polarization and back again,'' \emph{arXiv preprint arXiv:1908.09594}, 2019.

\bibitem{b36}
K.~Niu, K.~Chen, and J.-R. Lin, ``Beyond turbo codes: Rate-compatible punctured polar codes,'' in \emph{2013 IEEE International Conference on Communications (ICC)}, 2013, pp. 3423--3427.

\bibitem{b27}
V.~Bioglio, F.~Gabry, and I.~Land, ``Low-complexity puncturing and shortening of polar codes,'' in \emph{2017 IEEE Wireless Communications and Networking Conference Workshops (WCNCW)}, 2017, pp. 1--6.

\bibitem{b26}
R.~Wang and R.~Liu, ``A novel puncturing scheme for polar codes,'' \emph{IEEE Communications Letters}, vol.~18, no.~12, pp. 2081--2084, 2014.

\bibitem{b244}
X.~Gu, M.~Rowshan, and J.~Yuan, ``Rate-compatible shortened pac codes,'' in \emph{2023 IEEE/CIC International Conference on Communications in China (ICCC Workshops)}, 2023, pp. 1--6.

\bibitem{b4}
N.~Hussami, S.~B. Korada, and R.~Urbanke, ``Performance of polar codes for channel and source coding,'' in \emph{2009 IEEE International Symposium on Information Theory}, 2009, pp. 1488--1492.

\bibitem{b5}
M.~Bardet, V.~Dragoi, A.~Otmani, and J.-P. Tillich, ``Algebraic properties of polar codes from a new polynomial formalism,'' in \emph{2016 IEEE International Symposium on Information Theory (ISIT)}, 2016, pp. 230--234.

\bibitem{b6}
H.~Yao, A.~Fazeli, and A.~Vardy, ``A deterministic algorithm for computing the weight distribution of polar codes,'' in \emph{2021 IEEE International Symposium on Information Theory (ISIT)}, 2021, pp. 1218--1223.

\bibitem{li2021}
Y.~Li, H.~Zhang, R.~Li, J.~Wang, G.~Yan, and Z.~Ma, ``On the weight spectrum of pre-transformed polar codes,'' in \emph{2021 IEEE International Symposium on Information Theory (ISIT)}.\hskip 1em plus 0.5em minus 0.4em\relax IEEE, 2021, pp. 1224--1229.

\bibitem{rowshan2023}
M.~Rowshan, S.~H. Dau, and E.~Viterbo, ``On the formation of min-weight codewords of polar/pac codes and its applications,'' \emph{IEEE Transactions on Information Theory}, 2023.

\bibitem{rowshan2023closed}
M.~Rowshan, V.-F. Dr{\u{a}}goi, and J.~Yuan, ``On the closed-form weight enumeration of polar codes: 1.5$ d $-weight codewords,'' \emph{arXiv preprint arXiv:2305.02921}, 2023.

\bibitem{b17}
M.~Rowshan and J.~Yuan, ``Fast enumeration of minimum weight codewords of {PAC} codes,'' in \emph{2022 IEEE Information Theory Workshop (ITW)}, 2022, pp. 255--260.

\bibitem{zunker2024}
A.~Zunker, M.~Geiselhart, and S.~Ten~Brink, ``Enumeration of minimum weight codewords of pre-transformed polar codes by tree intersection,'' in \emph{2024 58th Annual Conference on Information Sciences and Systems (CISS)}.\hskip 1em plus 0.5em minus 0.4em\relax IEEE, 2024, pp. 1--6.

\bibitem{powerful}
V.~Miloslavskaya, B.~Vucetic, and Y.~Li, ``Computing the partial weight distribution of punctured, shortened, precoded polar codes,'' \emph{IEEE Transactions on Communications}, vol.~70, no.~11, pp. 7146--7159, 2022.

\bibitem{b23}
M.~Ellouze, R.~Tajan, C.~Leroux, C.~Jégo, and C.~Poulliat, ``Low-complexity algorithm for the minimum distance properties of pac codes,'' in \emph{2023 12th International Symposium on Topics in Coding (ISTC)}, 2023, pp. 1--5.

\bibitem{b19}
R.~Mori and T.~Tanaka, ``Performance and construction of polar codes on symmetric binary-input memoryless channels,'' in \emph{2009 IEEE International Symposium on Information Theory}, 2009, pp. 1496--1500.

\bibitem{b12}
{3GPP TS 38.212 V17.4.0}, ``{5G}; {NR}; multiplexing and channel coding,'' 2023.

\end{thebibliography}

\end{document}